\documentclass[journal]{IEEEtran}
\usepackage{booktabs}
\usepackage{amssymb}
\usepackage{color, cite}
\usepackage{mathrsfs}
\usepackage{graphicx}
\usepackage{amsmath,bm,amssymb}
\usepackage{multicol}
\usepackage{CJK}
\usepackage{indentfirst}
\usepackage{stfloats}
\usepackage{bm}
\usepackage{amsmath}

\usepackage{diagbox}
\usepackage{float}
\usepackage{multirow}
\usepackage{changepage}
\usepackage{amsfonts}
\usepackage{stfloats}
\usepackage{array}
\usepackage{titletoc}
\usepackage{algorithm}
\usepackage{algorithmic}
\usepackage{caption}
\usepackage{subfigure}  
\usepackage[font={small}]{caption}
\usepackage{mdframed}



\begin{document}

\title{ High-Efficient Near-Field Channel Characteristics Analysis for Large-Scale MIMO Communication Systems }

\author{Hao Jiang, \IEEEmembership{Member, IEEE}, Wangqi Shi, Xiao Chen, \IEEEmembership{Member, IEEE},  Qiuming Zhu, \IEEEmembership{Senior Member, IEEE},\\ and Zhen Chen, \IEEEmembership{Senior Member, IEEE}
\thanks{This work was supported by the National Natural Science Foundation of China (NSFC) projects (No. 62101275 and 62101273). }
\thanks{H. Jiang, W. Shi, and X. Chen are with the School of Artificial Intelligence, Nanjing University of Information Science and Technology, Nanjing 210044, P. R. China. H. Jiang is also with the National Mobile Communications Research Laboratory, Southeast University, Nanjing 210096, P. R. China (emails: jianghao@nuist.edu.cn; shiwangqi@nuist.edu.cn; x.chen@nuist.edu.cn).}
\thanks{Q. Zhu is with the College of Electronic and Information Engineering, Nanjing University of Aeronautics and Astronautics, Nanjing 211106, China (e-mail: zhuqiuming@nuaa.edu.cn). }
\thanks{Z. Chen is with the Department of Electrical Engineering, City University of Hong Kong, Hong Kong  (e-mail: chenz.scut@gmail.com).}

}

\maketitle

\begin{abstract}

Large-scale multiple-input multiple-output (MIMO) holds great promise for the fifth-generation (5G) and future communication systems. In near-field scenarios, the spherical wavefront model is commonly utilized to accurately depict the propagation characteristics of large-scale MIMO communication channels. However, employing this modeling method necessitates the computation of angle and distance parameters for each antenna element, resulting in challenges regarding computational complexity. To solve this problem, we introduce a subarray decomposition scheme with the purpose of dividing the whole large-scale antenna array into several smaller subarrays. This scheme is implemented in the near-field channel modeling for large-scale MIMO communications between the base stations (BS) and the mobile receiver (MR). Essential channel propagation statistics, such as spatial cross-correlation functions (CCFs), temporal auto-correlation functions (ACFs), frequency correlation functions (CFs), and channel capacities, are derived and discussed. A comprehensive analysis is conducted to investigate the influence of the height of the BS, motion characteristics of the MR, and antenna configurations on the channel statistics. The proposed channel model criterions, such as the modeling precision and computational complexity, are also theoretically compared. Numerical results demonstrate the effectiveness of the presented communication model in obtaining a good tradeoff between modeling precision and computational complexity.

\end{abstract}

\IEEEpeerreviewmaketitle

\section{Introduction}

\subsection{Background}
{\color{blue}
Sixth generation (6G) wireless communication systems aim to establish a multi-dimensional information network of space-air-ground-sea, providing users with extremely low latency and high data rate services.} This will put forward higher requirements for the performance and spectrum of mobile communication technology \cite{Guig,Tangfx}. As an emerging technology, large-scale multiple-input multiple-output (MIMO) holds the promise of supporting huge throughput, massive connectivity, and improved energy efficiency \cite{Ruan2024,ChenZ2024}. Deploying large-scale MIMO on base stations (BS) enables effective interference suppression among users and provides favorable propagation conditions \cite{Yuan2023,Zhuy}. As a result, large-scale MIMO communication is regarded as an important direction of 6G and its development is closely monitored. The introduction of large-scale MIMO technologies also brings about complexities in channel characteristics, including propagation non-stationarities, near-field conditions, and spherical wavefronts. This requires deeper researches through accurate channel modeling \cite{Jiang2019MIMO, Wei2022, ShiW2024}.

\begin{figure}
  \centering
  \includegraphics[width=8.8cm]{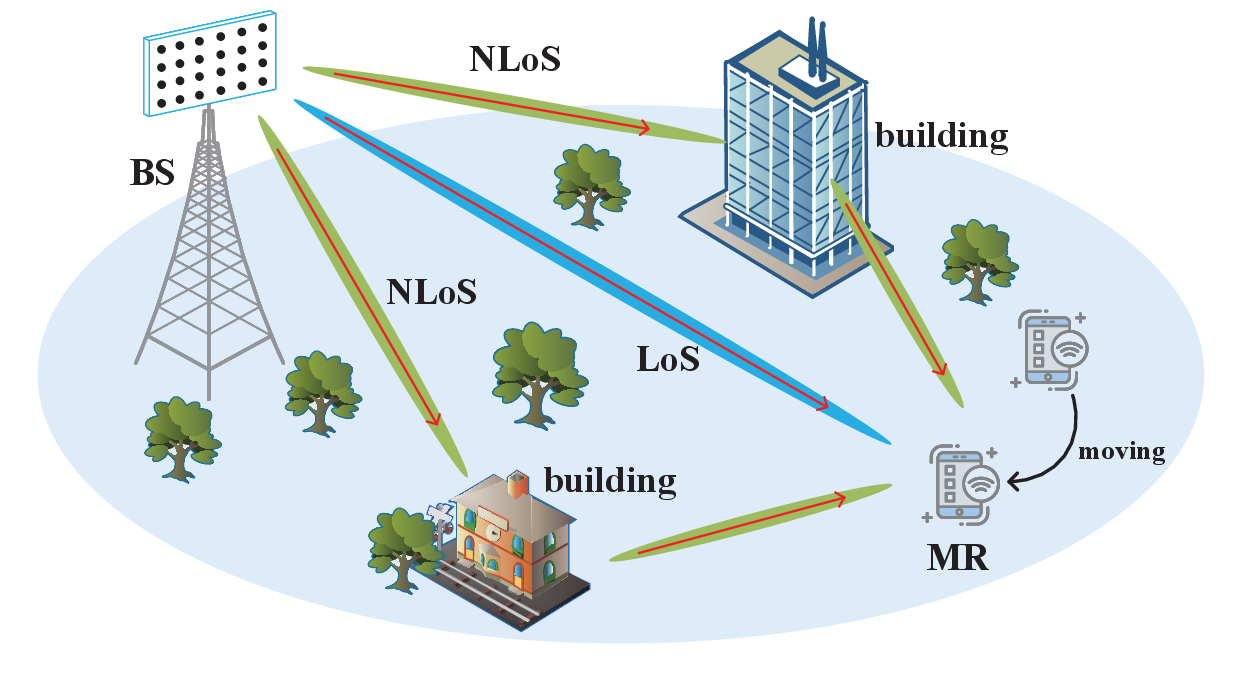}\\
  \caption{Illustration of BS-MR communication in the large-scale MIMO channel model.}
  \label{fig0}
\end{figure}

\subsection{Related Works}

Channel models play crucial roles in accurately describing the transmission characteristics for large-scale MIMO communications, making them essential for the design and optimization of wireless communication systems \cite{Mao2024, Jiang2018, Bai2024}. In \cite{JiangV2V2020}, the authors establish a three-dimensional (3D) channel model based on the geometric stochastic model and analyze the channel non-stationarity within large-scale MIMO communication systems, including space and time domains. The authors in \cite{Yang2024} propose a non-stationary channel model for large-scale MIMO communication systems and investigate the channel statistical characteristics and channel capacity. In \cite{Jiang2021}, the authors present a pervasive wireless channel modeling theory, which possesses the capability to describe the channel statistical characteristics for large-scale MIMO communication systems. However, the significant increase in antenna array size brings changes in channel characteristics, which is particularly evident in near-field effects \cite{Liu2023, JWang2024}.
{\color{blue}
In spatial domain, the near-field and far-field ranges are distinguished by the Rayleigh distance, which can be calculated as $2 D^2/\lambda$ with $D$ and $\lambda$ being the antenna apertures and the wavelength, respectively \cite{Han2024}.} Specifically, when the path length connecting the transmitter and receiver is less than the Rayleigh distance, the large-scale MIMO channel model typically assumes a spherical wavefront, while the planar wavefront assumption is utilized when the distance exceeds the Rayleigh distance \cite{Lu20231}. This allows for highly accurate large-scale MIMO channel modeling. In the previous channel modeling works, the Rayleigh distance typically remain very small due to the limitation of the number of antennas, leading to its frequent omission in the modeling process. Nevertheless, the applications of large-scale MIMO have significantly widened the near-field range, thereby the near-field effect cannot be overlooked in current channel modeling \cite{He2023}. The authors in \cite{Wang2024near} present a near-field communication model for large-scale MIMO systems by employing the assumption of non-uniform spherical waves, which serves as an innovative substitution for the conventional plane-wavefront assumption. In \cite{Chen2022}, the authors develop a channel model that combines spherical and planar waves to evaluate the performance in both near- and far-field scenarios, demonstrating the modeling ability to accurately describe the hybrid characteristics of these regions. In practice, the Rayleigh distance in large-scale MIMO systems is often greater than the path length connecting the transmitter and the receiver, making it necessary to focus on the near-field channel modeling \cite{Zhang2023}.

However, the computational complexity associated with the near-field spherical wavefront model cannot be ignored, as it necessitates the calculation of angle and distance parameters for each antenna element. Especially when the antenna array expands, the dimension of channel matrix will increase sharply, which poses a great challenge to channel modeling \cite{RuanC2024, Huang2024}. To address this issue, the authors in \cite{Jiang2023} introduce the subarray decomposition framework for the reconfigurable intelligence surface (RIS)-empowered communication systems. By decomposing the RIS array, the channel model in \cite{Jiang2023} can effectively convert the near-field communication scenario into a far-field one, thus reducing the computational complexity. Given the considerable complexity of the spherical wavefront model in near-field ranges, this approach holds the potential to simplify modeling in large-scale MIMO communication systems. Similarly, in \cite{Cui2021}, the authors present the subarray decomposition scheme to deal with the near-field beam splitting effect. In \cite{Ribeiro2021}, the authors incorporate vertical subarrays into precoding schemes, aiming at reducing the complexity without losing much performance. Moreover, the subarray decomposition scheme is also mentioned in \cite{Lu2023}, which aims to reduce hardware cost and energy consumption for extremely large antenna array systems.
{\color{blue}
However, the subarray decomposition scheme proposed in \cite{Jiang2023} mainly concentrate on the performance optimization of RIS-assisted communication systems, while the primary objectives of \cite{Cui2021, Ribeiro2021, Lu2023} are related to the signal processing and hardware costs, which are significantly different from our research work. Furthermore, the authors in \cite{ChangBDCM2023} partition a large-scale antenna array into multiple smaller subarrays for distinguishing the near- and far-field ranges; however, they fail to delve into the implications of this method on the computational complexity. Our study, in contrast, aims to address this gap by comprehensively evaluating the effect of subarray partitioning on both system performance and computational complexity.}

For the sake of fulfilling the technical requirements for future communication systems, such as RIS and THz communication, a large number of antennas is necessary to attain the desired performance. Consequently, the investigation of large-scale MIMO systems holds great importance \cite{Cui2023, Li2022}. However, this results in an exponential growth in the Rayleigh distance of the antenna array, leading to fall in near-field propagation scenarios and invalidating the traditional planar wavefront assumption.
{\color{blue}
Most of the existing works on large-scale MIMO channel modeling have not discussed the modeling differences between the near-field and far-field ranges; meanwhile, the existing ones related to the subarray method have not achieved the optimal balance between the channel modeling precision and computational complexity for MIMO communication systems. In this case, if we continue to use the planar wavefront model to study the large-scale MIMO communication systems performance, the modeling precision will fall far from the required criterion. Furthermore, wireless channel modeling based on the spherical wavefront assumption would lead to significant computational modeling complexity, which results in high hardwave overhead for computer systems. Therefore, the main research goal of this paper is to develop near-field channel model for large-scale MIMO communication systems with the help of the subarray decomposition scheme, which aims at reducing the modeling complexity under the premise of ensuring modeling precision.}

\subsection{Main Contributions}

\begin{figure}
  \centering
  \includegraphics[width=8.8cm]{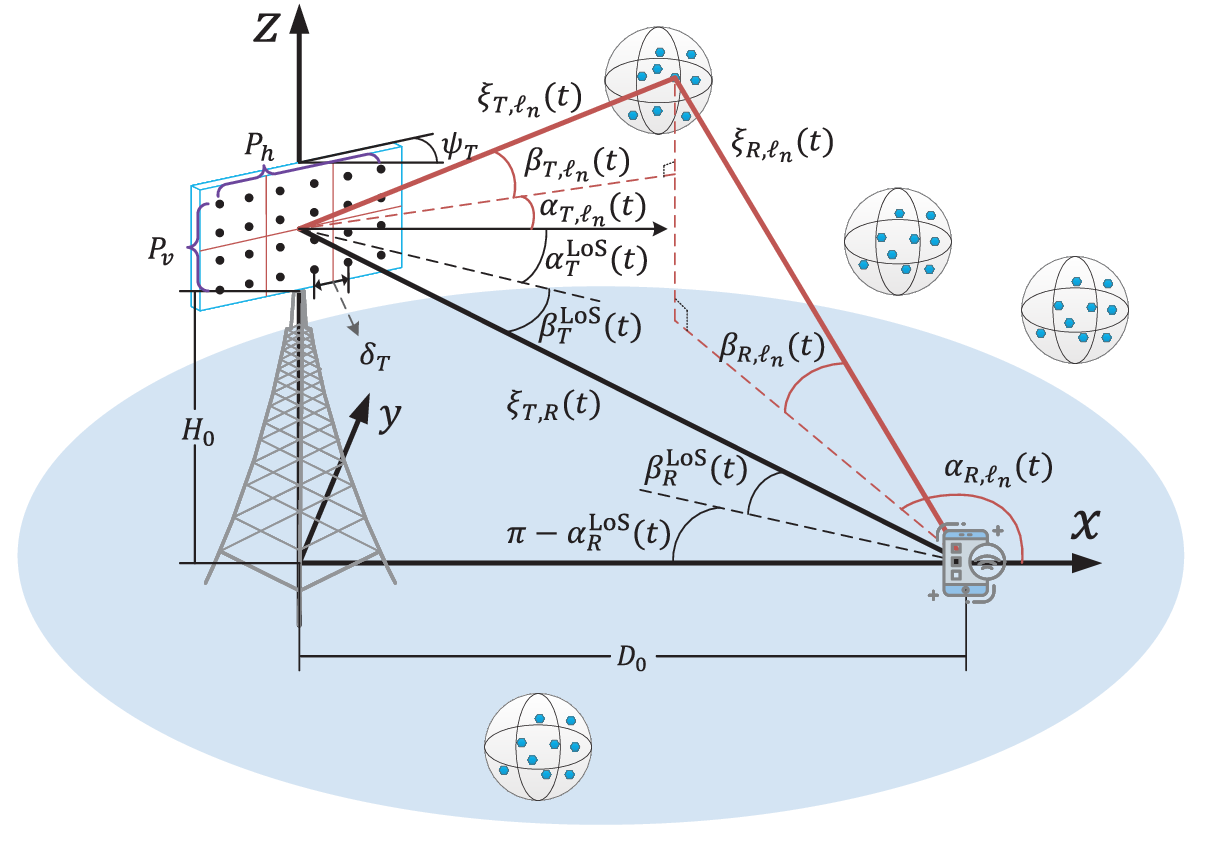}\\
  \caption{Illustration of angle and path parameters in the large-scale MIMO channel model for near-field scenarios.}
  \label{fig1}
\end{figure}

In this article, we propose a 3D non-stationary channel model for large-scale MIMO systems with the purpose of effectively capturing the channel characteristics in near-field communication environments. The proposed model decomposes the large-scale transmit antenna array into several smaller units, which aims at achieving the technical goal of reducing the computational complexity. {\color{blue} Comparative results demonstrate that the proposed channel modeling approach exhibits high precision with low complexity.}

\begin{figure}[!t]
  \centering
  \includegraphics[width=8.8cm]{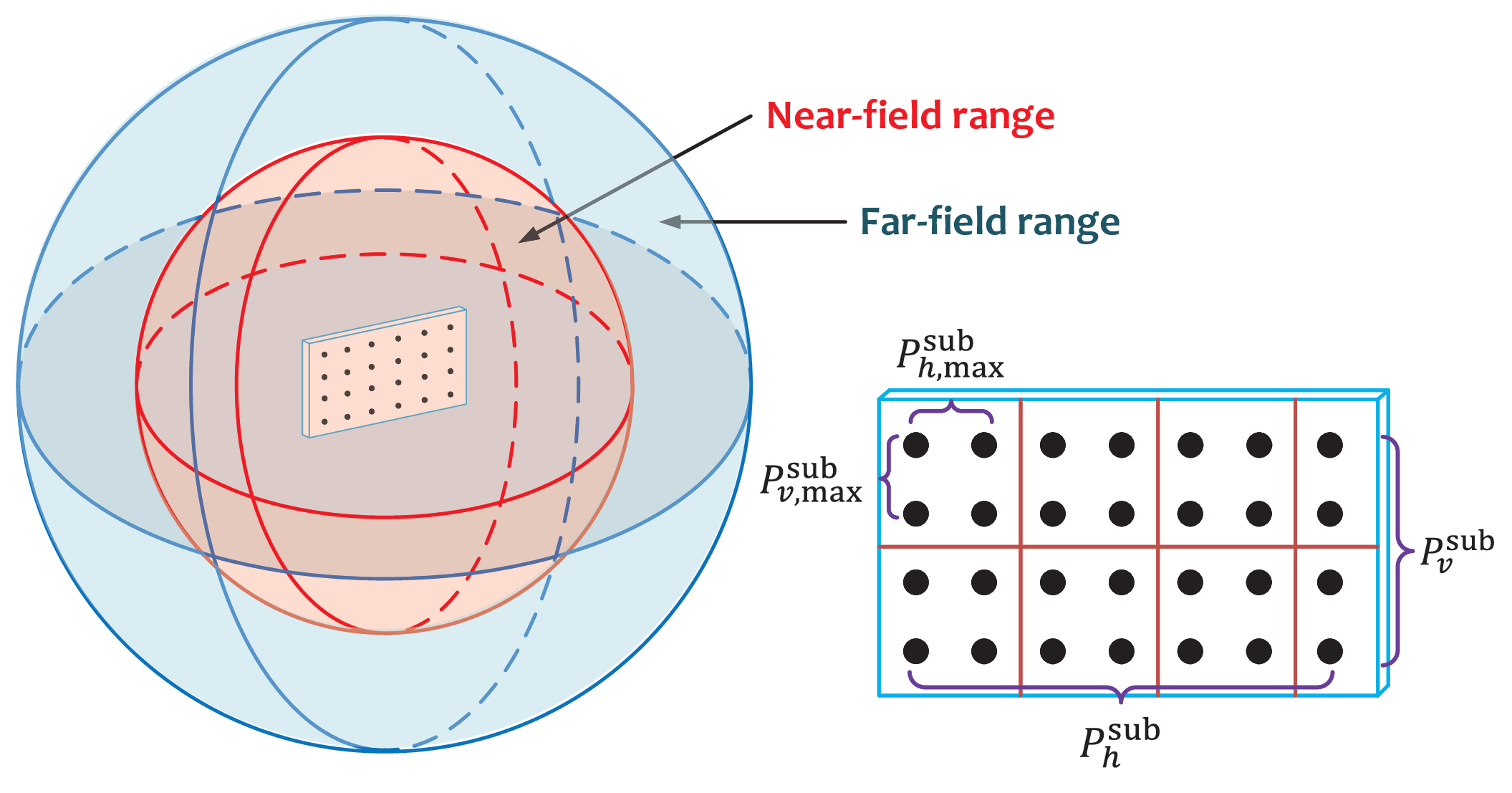}\\
  \caption{Illustration of near- and far-field ranges in large-scale MIMO channel model and the proposed subarray decomposition scheme.}
  \label{fig2}
\end{figure}

{\color{blue}
The main contributions are summarized as follows:

\begin{itemize}
 \item  We propose a subarray decomposition framework to address the problem of high computational complexity for large-scale MIMO communication systems. This framework involves decomposing the large-scale antenna array at the BS side, ensuring that each decomposed subarray satisfies the planar wavefront assumption. We verify the precision criterion of the framework in investigating the channel characterization within the spatial, time, and frequency domains.
\end{itemize}

\begin{itemize}
 \item  We undertake a comparative analysis of the modeling performance between the designed subarray decomposition scheme and the conventional planar/spherical wavefront assumptions, while also demonstrating the low computational complexity of the channel modeling. This assessment highlights the designed subarray decomposition framework in achieving the high-precision modeling while maintaining low computational complexity.
\end{itemize}

\begin{itemize}
 \item 	We utilize the subarray decomposition framework to achieve the optimal balance between the modeling precision and computational complexity for large-scale MIMO channel modeling. Essential channel statistical properties, such as spatial cross-correlation functions (CCFs), temporal auto-correlation functions (ACFs), frequency correlation functions (CFs), and channel capacities are also derived and discussed. In addition, we explore the influences of parameters such as time interval and motion attributes of the receiver on channel propagation characteristics. These observations provide valuable insights and references for the performance evaluation and optimization of large-scale MIMO systems.
\end{itemize}
}

The remainder of this paper is summarized below. In Section II, we provide the communication model based upon the subarray decomposition scheme. Section III derives the propagation properties for the proposed channel model. The simulation results and conclusions of the paper are given in Sections IV and V, respectively.

\textbf{Notation}: The lowercase letters (e.g., $x$), boldface lowercase letters (e.g., $\textbf{x}$), and boldface uppercase letters (e.g., $\textbf{X}$) respectively represent the scalars, vectors, and matries. $\Vert \cdot \Vert$, ($\cdot$)$^*$, $[ \hspace*{0.025cm} \cdot \hspace*{0.025cm} ]^\text{T}$ stand for the Frobenius norm, complex conjugate
operation, and transpose operation of matries, respectively.
{\color{blue}
Lastly, $j = \sqrt{-1}$ represents the imaginary unit and $\mathbb{E}[\cdot]$ denotes the operation of expectation.}

\section{System Model}

{\color{blue}
As shown in Figs. \ref{fig0} and \ref{fig1}, we consider a physics-based channel model in 3D free space for BS-to-mobile receiver (MR) communications in near-field scenarios, where $P_h \times P_v$ omni-directional uniform planar array (UPA) antennas are equipped at the BS side, while $Q$ omni-directional uniform linear array (ULA) antennas are equipped at the MR side \cite{LiuT2023}. It is worth mentioning that the proposed channel model is operated at the sub-6G frequency band, with a bandwidth of $50$ MHz, which is dedicated for large-scale MIMO communication systems \cite{Cui2021}.} The height of the BS is expressed as $H_0$, and the distance from the MR to the bottom midpoint of the BS is $D_0$. Taking the projection of the midpoint of the BS antenna array on the horizontal plane as the origin of the global coordinate system, we define the positive direction of the $x$-axis as the line connecting the origin coordinate and the midpoint of the antenna array at the receiver side. The $z$-axis goes straight up through the origin, and the $y$-axis is determined in accordance with the right-hand rule. Additionally, the path length vector from the origin coordinate to the midpoints of the antenna arrays at the BS and MR sides are denoted by $\mathbf{d}_T = [ \, 0, 0, H_0 + 0.5P_v \delta_T \,]^\mathrm{T}$ and $\mathbf{d}_R = [\, D_0, 0, 0 \,]^\mathrm{T}$, respectively, with $\delta_T$ being the spacing between two neighboring antennas in BS UPA. Therefore, we can obtain the distance vector of the $q$-th $(q = 1, 2,..., Q)$ receiving antenna regarding the coordinate origin as follows:
\begin{eqnarray}
\mathbf{d}_q (t) \hspace*{-0.225cm}&=&\hspace*{-0.225cm} \left [ \begin{array}{ccc} \, d_{q,x} (t) \, \\[0.125cm] d_{q,y} (t) \\[0.125cm] d_{q,z} \end{array} \right ] \nonumber \\[0.125cm]
\hspace*{-0.225cm}&=&\hspace*{-0.225cm} \left [ \begin{array}{ccc} \, D_0 + k_q \delta_R \cos\psi_R \cos{\theta_R} + v_R t \cos{\eta_R} \, \\[0.125cm] k_q \delta_R \sin\psi_R \cos{\theta_R} + v_R t \sin{\eta_R} \\[0.125cm] k_q \delta_R \sin{\theta_R} \end{array} \right ] \, ,
\end{eqnarray}
where $k_q = (Q - 2q + 1)/2$ and $\delta_R$ denotes the spacing between two neighboring antennas in MR ULA. To further enhance the universality capabilities of the provided channel model, let $\psi_T$ and $\psi_R$ represent the azimuth orientation angles that characterize the antenna arrays positioned at the BS and MR sides, respectively, so that the position layout of multiple antenna arrays can be simulated. Notably, when the length between the BS and MR is less than the Rayleigh distance, which corresponds to the near-filed scenarios in Fig.~\ref{fig2}, the spherical wavefront is considered to characterize the channel characteristics. However, when the BS-to-MR length exceeds the Rayleigh distance, the planar wavefront should be taken into account for this far-field communication modeling. The Rayleigh distance, which distinguishes the near- and far-field ranges, can be expressed as
\begin{eqnarray}\label{eq:L}
L = \frac{2 d^2_{dia}}{\lambda} = \frac{2 \delta_T^2 \big((P_h-1)^2 + (P_v-1)^2 \big)}{\lambda} \, ,
\end{eqnarray}
where $d_{dia}$ denotes the diagonal length of the antenna array. It is worth emphasizing that our division of the near- and far-fields, based upon the Rayleigh distance, accurately reflects the physical phenomena within the current simulation parameters, while also acknowledging the influence of angular parameters and their importance in future considerations. The Rayleigh distances measured in meters under different array apertures are shown in Table \ref{table1}. Notably, an enlargement in the aperture size of the antenna array at the BS side leads to a corresponding increase in the Rayleigh distance within the channel model. Specifically, an antenna array with size $2 \, \text{m} \times 2 \, \text{m}$ has a Rayleigh distance of $128$ m at $2.4$ GHz band and $267$ m at $5$ GHz band, which is significantly larger than the radius of the typical fifth-generation (5G) cell \cite{Cui2023}. For the near-field channel modeling with a large number of antennas, the subarray decomposition scheme has the potential to reduce the computational complexity.

\begin{table}
    \centering
    \fontsize{8}{14}\selectfont    
    \caption{Illustration of antenna array Rayleigh distances.}
\begin{tabular}{|c|c|c|c|c|}
\hline
  \diagbox{Frequency}{Size}& $1 \, \text{m} \times 0.1 \, \text{m}$ & $1 \, \text{m} \times 2 \, \text{m}$  & $2 \, \text{m} \times 2 \, \text{m}$\\
\hline
$2.4 \, \text{GHz}$ & $16$ m & $80$ m & $128$ m\\
\hline
$5 \, \text{GHz}$ & $34$ m & $167$ m & $267$ m\\
\hline
\end{tabular}
\label{table1}
\end{table}

\subsection{Proposed Subarray Decomposition Scheme}

{\color{blue}
In the existing literature, such as \cite{Yuan2023} and \cite{Jiang2023}, channel modeling for large-scale MIMO communications in near-field ranges primarily relies on the spherical wavefront model to capture the channel propagation characteristics. Although this approach offers high modeling precision criterion, its computational complexity cannot be overlooked due to the requirement of calculating angle and distance parameters for each antenna. Especially for the case of large antenna arrays, the computational complexity will be extremely high and cannot be acceptable. Furthermore, owing to the fact that the planar wavefront model does not take into account the deviation of the signal angles in the antenna array; therefore, the investigation of the propagation statistics based on the planar wavefront model cannot achieve the sufficient modeling accuracy. In light of this, it is of vital importance to design a scheme that effectively obtain the optimal balance between the precision and complexity for large-scale MIMO communication channel modeling.} To deal with this challenge, we present a subarray decomposition scheme, as illustrated in Fig. \ref{fig2}. This scheme divides whole transmitting antenna array into $P^{\text{sub}}_h \times P^{\text{sub}}_v$ units, with the dimension of the largest unit being $P^{\text{sub}}_{h,\text{max}} \times P^{\text{sub}}_{v,\text{max}}$. For each subarray, all of its elements share the location coordinates of the center point, ensuring that antennas within the same subarray exhibit consistent angle and distance parameters.
Therefore, the $P^{\text{sub}}_h$ and $P^{\text{sub}}_v$ can be respectively derived by
\begin{eqnarray}
\hspace*{-0.5cm} P^{\text{sub}}_h \hspace*{-0.225cm}&=&\hspace*{-0.225cm} \left \{
                \begin{array}{lr}
                \frac{P_h - \bmod(P_h, P^{\text{sub}}_{h,\text{max}})}{P^{\text{sub}}_{h,\text{max}}} + 1, \\ [0.125cm]
                \hspace{2.4cm}\text{if} \bmod(P_h, P^{\text{sub}}_{h,\text{max}}) \neq 0 \\ [0.125cm]
                \frac{P_h}{P^{\text{sub}}_{h,\text{max}}}, \hspace{1.435cm}\text{if} \bmod(P_h, P^{\text{sub}}_{h,\text{max}}) = 0 \\
                \end{array}
            \right . \, ,
\end{eqnarray}
\begin{eqnarray}
\hspace*{-0.5cm} P^{\text{sub}}_v \hspace*{-0.225cm}&=&\hspace*{-0.225cm} \left \{
                \begin{array}{lr}
                \frac{P_v - \bmod(P_v, P^{\text{sub}}_{v,\text{max}})}{P^{\text{sub}}_{v,\text{max}}} + 1, \\ [0.125cm]
                \hspace{2.4cm}\text{if} \bmod(P_v, P^{\text{sub}}_{v,\text{max}}) \neq 0 \\ [0.125cm]
                \frac{P_v}{P^{\text{sub}}_{v,\text{max}}}, \hspace{1.445cm}\text{if} \bmod(P_v, P^{\text{sub}}_{v,\text{max}}) = 0 \\
                \end{array}
            \right . \, .
\end{eqnarray}
{\color{blue}
It is important to note that the size of any subarray is limited by the dimension of the entire array, i.e., $P^{\text{sub}}_{h,\text{max}} < P_h$ and $P^{\text{sub}}_{v,\text{max}} < P_v$, which indicates that the dimension of the $(p^{\text{sub}}_h, p^{\text{sub}}_v)$-th $(p^{\text{sub}}_h = 1,2,...,P^{\text{sub}}_h$ and $p^{\text{sub}}_v = 1,2,...,P^{\text{sub}}_v)$ subarray in the transmitting antenna array can be expressed as}
\begin{eqnarray}
\hspace*{-0.5cm} P_{p^{\text{sub}}_h} \hspace*{-0.225cm}&=&\hspace*{-0.225cm} \left \{
                    \begin{array}{lr}
                    P^{\text{sub}}_{h,\text{max}}, \hspace{1.73cm}\text{if} ~ 1 \leq p^{\text{sub}}_h < P^{\text{sub}}_h \\ [0.125cm]
                    P_h - (P^{\text{sub}}_h - 1)P^{\text{sub}}_{h,\text{max}},
                    \hspace{0.4cm}\text{if} ~ p^{\text{sub}}_h = P^{\text{sub}}_h \\
                    \end{array}
                \right . \, ,
\end{eqnarray}
\begin{eqnarray}
\hspace*{-0.5cm} P_{p^{\text{sub}}_v} \hspace*{-0.225cm}&=&\hspace*{-0.225cm} \left \{
                    \begin{array}{lr}
                    P^{\text{sub}}_{v,\text{max}}, \hspace{1.745cm}\text{if} ~ 1 \leq p^{\text{sub}}_v < P^{\text{sub}}_v \\ [0.125cm]
                    P_v - (P^{\text{sub}}_v - 1)P^{\text{sub}}_{v,\text{max}},
                    \hspace{0.4cm}\text{if} ~ p^{\text{sub}}_v = P^{\text{sub}}_v \\
                    \end{array}
                \right . \, .
\end{eqnarray}
Therefore, the path length vector connecting the origin coordinate and the midpoint of the $(p^{\text{sub}}_h, p^{\text{sub}}_v)$-th subarray in the transmitting antenna array can be derived by
\begin{align}
&\mathbf{d}_{p^{\text{sub}}_{h,v}} =
                \left [
                    \begin{array}{ccc}
                    d_{p^{\text{sub}}_{h,v},x} \\ [0.125cm]
                    d_{p^{\text{sub}}_{h,v},y} \\ [0.125cm]
                    d_{p^{\text{sub}}_{h,v},z}
                    \end{array}
                \right ] \nonumber \\[0.125cm]
                &= \left [
                       \begin{array}{ccc}
                       \big( (p^{\text{sub}}_h - 1)P^{\text{sub}}_{h,\text{max}} + 0.5 P_{p^{\text{sub}}_h} - 0.5 P_h \big) \delta_T \cos{\psi_T} \\ [0.125cm]
                       \big( (p^{\text{sub}}_h - 1)P^{\text{sub}}_{h,\text{max}} + 0.5 P_{p^{\text{sub}}_h} - 0.5 P_h \big) \delta_T \sin{\psi_T} \\ [0.125cm]
                       H_0 + \big( (p^{\text{sub}}_v - 1)P^{\text{sub}}_{v,\text{max}} + 0.5 P_{p^{\text{sub}}_v} \big)\delta_T
                       \end{array}
                   \right ] \, .
\end{align}
{\color{blue}
In the UPA of the BS side, the subarray containing the $(p_h, p_v)$-th element, that is, $(p^{\text{sub}}_h, p^{\text{sub}}_v)$-th subarray, can be expressed as \cite{Jiang2023}}
\allowdisplaybreaks[3]
\begin{eqnarray}
\hspace*{-0.5cm} p^{\text{sub}}_h \hspace*{-0.225cm}&=&\hspace*{-0.225cm} \left \{
                \begin{array}{lr}
                \frac{p_h - \bmod(p_h, P^{\text{sub}}_{h,\text{max}})}{P^{\text{sub}}_{h,\text{max}}} + 1, \\ [0.125cm]
                \hspace{2.4cm}\text{if} \bmod(p_h, P^{\text{sub}}_{h,\text{max}}) \neq 0 \\ [0.125cm]
                \frac{p_h}{P^{\text{sub}}_{h,\text{max}}}, \hspace{1.435cm}\text{if} \bmod(p_h, P^{\text{sub}}_{h,\text{max}}) = 0 \\
                \end{array}
            \right . \, ,
\end{eqnarray}
\begin{eqnarray}
\hspace*{-0.5cm} p^{\text{sub}}_v \hspace*{-0.225cm}&=&\hspace*{-0.225cm} \left \{
                \begin{array}{lr}
                \frac{p_v - \bmod(p_v, P^{\text{sub}}_{v,\text{max}})}{P^{\text{sub}}_{v,\text{max}}} + 1, \\ [0.125cm]
                \hspace{2.4cm}\text{if} \bmod(p_v, P^{\text{sub}}_{v,\text{max}}) \neq 0 \\ [0.125cm]
                \frac{p_v}{P^{\text{sub}}_{v,\text{max}}}, \hspace{1.445cm}\text{if} \bmod(p_v, P^{\text{sub}}_{v,\text{max}}) = 0 \\
                \end{array}
            \right . \, .
\end{eqnarray}
It shows that the choices of the size setting of the largest subarray in the subarray decomposition scheme for large-scale MIMO systems have direct impacts on both modeling precision and complexity criterions. When we expand the size of the largest unit, the modeling precision and complexity will correspondingly decline, and vice versa. We regard that when MR is in the far-field range and close to the boundary of near- and far-field, the subarray decomposition reaches the optimal size.
{\color{blue}
It is worth mentioning that the proposed subarray decomposition scheme makes it possible to realize all the subarrays to satisfy the far-field planar wavefront assumption, which leads to a phenomenon that all the elements in the same subarray share the same distance/angle parameters. Compared with the conventional spherical wavefront model, the proposed subarray decomposition scheme has great potential in reducing the modeling computational complexity, while ensuring modeling accuracy, thereby playing great roles in the performance evaluation and analysis for large-scale MIMO communication systems.}

\subsection{Complex Channel Impulse Response (CIR)}

For the presented communication model, the waves transmitted from the BS undergo two distinct kinds of propagation components before reaching the MR, they are, the direct line-of-sight (LoS) component and the non-line-of-sight (NLoS) component.
{\color{blue}
We investigate the physical properties of the proposed channel model through an matrix $\mathbf{H} (t,\tau) = [ \, h_{pq} (t,\tau) \, ]_{Q \times P_h P_v}$, with $\tau$ being the path delay}. Here, the $p$-th transmitting element is located at the $p_h$-th $(p_h = 1,2,...,P_h)$ row and $p_v$-th $(p_v = 1,2,...,P_v)$ column. The $h_{pq} (t,\tau)$, which indicates the complex channel impulse response (CIR) from the $p$-th element at the BS side to the $q$-th element at the MR side. Assume that these two kinds of propagation components operate independently of each other, thus the $h_{pq} (t,\tau)$ can be derived by the sum of the CIRs of the LoS and NLoS paths, that is \cite{ZengL2024}
\begin{eqnarray}\label{eq:h_pq}
h_{pq} (t,\tau) \hspace*{-0.225cm}&=&\hspace*{-0.225cm} \sqrt{\frac{K}{K+1}} h^{\text{LoS}}_{pq} (t) \delta \big( \tau - \tau^{\text{LoS}}(t) \big) \nonumber \\[0.125cm]
\hspace*{-0.225cm}&+&\hspace*{-0.225cm} \sqrt{\frac{1}{K+1}} h^{\text{NLoS}}_{pq} (t) \delta \big( \tau - \tau^{\text{NLoS}}(t) \big) \,  ,
\end{eqnarray}
with $K$ being the Rican factor. $\tau^{\text{LoS}}(t) = \xi_{T,R}(t)/c$ and $\tau^{\text{NLoS}}(t) = (\xi_{T,\ell_n} + \xi_{R,\ell_n} (t))/c$, where $c$ is the speed of light. Here, $\xi_{T,R}(t) = \Vert \mathbf{d}_R (t) - \mathbf{d}_T \Vert$ denotes the propagation path length connecting the midpoints of the antenna arrays at the BS and MR sides; $\xi_{T,\ell_n} = \Vert \mathbf{d}_{\ell_n} - \mathbf{d}_T \Vert$ and $\xi_{R,\ell_n} (t) = \Vert \mathbf{d}_{\ell_n} - \mathbf{d}_R (t) \Vert$ are respectively the propagated distances of the waves connecting the $n$-th path in the $\ell$-th cluster and the midpoints of the antenna arrays at the BS and MR sides. In \eqref{eq:h_pq}, $h^\text{LoS}_{pq} (t)$ is the channel coefficient within the LoS propagation component for the $(p,q)$-th antenna pair, which expression can be shown as follows:
\allowdisplaybreaks[3]
\begin{eqnarray}\label{eq:h_LoS}
h^{\text{LoS}}_{pq} (t) \hspace*{-0.225cm}&=&\hspace*{-0.225cm} e^{-j \frac{2 \pi}{\lambda} \xi_{T,R} (t) } \nonumber \\[0.125cm]
\hspace*{-0.225cm}&\times&\hspace*{-0.225cm} e^{j \frac{2\pi}{\lambda} k_{p_h} \delta_T \cos \big(\alpha^{\text{LoS}}_{T}(t) - \psi_T \big) \cos\beta^{\text{LoS}}_{T}(t) } \nonumber \\[0.125cm]
\hspace*{-0.225cm}&\times&\hspace*{-0.225cm} e^{j \frac{2\pi}{\lambda} k_{p_v} \delta_T \sin\beta^{\text{LoS}}_{T}(t) } \nonumber \\[0.125cm]
\hspace*{-0.225cm}&\times&\hspace*{-0.225cm} e^{j \frac{2\pi}{\lambda} k_q \delta_R \cos \big(\alpha^{\text{LoS}}_{R}(t) - \psi_R \big) \cos\beta^{\text{LoS}}_{R}(t) \cos{\theta_R} } \nonumber \\[0.125cm]
\hspace*{-0.225cm}&\times&\hspace*{-0.225cm}  e^{j \frac{2\pi}{\lambda} k_q \delta_R \sin\beta^{\text{LoS}}_{R}(t) \sin{\theta_R} } \nonumber \\[0.125cm]
\hspace*{-0.225cm}&\times&\hspace*{-0.225cm}  e^{j \frac{2\pi}{\lambda} v_R t \cos \big(\alpha^{\text{LoS}}_{R}(t) - \eta_R \big) \cos{\beta^{\text{LoS}}_{R}(t)} }\, ,
\end{eqnarray}
where $k_{p_h} = (P_h - 2p_h + 1)/2$ and $k_{p_v} = (P_v - 2p_v + 1)/2$. The $\alpha^\text{LoS}_T (t)$ and $\beta^\text{LoS}_T (t)$ denote the angles of departure of the transmitted waves in the azimuth and vertical planes, respectively, which can be derived by calculating the angles between the ray of the LoS propagation link and the $x$-axis.
As a result, we have
\begin{eqnarray}
\hspace*{-1.5cm} \alpha^{\text{LoS}}_{T}(t) \hspace*{-0.225cm}&=&\hspace*{-0.225cm} \arctan\frac{d_{q,y}(t) - d_{p^{\text{sub}}_{h,v},y}}{d_{q,x}(t) - d_{p^{\text{sub}}_{h,v},x}} \, ,
\end{eqnarray}
\begin{eqnarray}
\hspace*{-1.5cm} \beta^{\text{LoS}}_{T}(t) \hspace*{-0.225cm}&=&\hspace*{-0.225cm} \nonumber\\
&&\hspace*{-1.6cm} \arctan\frac{d_{p^{\text{sub}}_{h,v},z} - d_{q,z}}{ \sqrt{(d_{q,x}(t) - d_{p^{\text{sub}}_{h,v},x})^2 + (d_{q,y}(t) - d_{p^{\text{sub}}_{h,v},y})^2} } \, .
\end{eqnarray}
Based on the geometry properties, the angle of arrival of the waves at the MR side in the azimuth and vertical planes are derived as $\alpha^{\text{LoS}}_{R} (t) = \pi - \alpha^{\text{LoS}}_{T} (t)$ and $\beta^{\text{LoS}}_{R} (t) = \beta^{\text{LoS}}_{T} (t)$, respectively. Obviously, Eq. (11) consists of the distance and angle parameters at the BS and MR sides, which effectively captures the physical characteristics of the wireless communication scenario.

For the NLoS component, we define the distance vector from the origin coordinate to the $n$-th $(n = 1, 2, ...,\ell_N)$ path in $\ell$-th $(\ell = 1, 2, ...,L)$ cluster as $\mathbf{d}_{\ell_n} = [ \, x_{\ell_n}, y_{\ell_n}, z_{\ell_n} \, ]^{\mathrm{T}}$. It is worth mentioning that we only focus on the analysis of small-scale fading characteristics, while large-scale fading characteristics has been researched and discussed in our previous works \cite{JiangRIS2023, Xiong2022UAV}, thus large obstacles are not considered in the proposed channel model. The channel coefficient within the NLoS link for the $(p, q)$-th antenna pair, i.e., $h^{\text{NLoS}}_{pq} (t)$, is derived by
\begin{eqnarray}\label{eq:h_NLoS}
\hspace*{-0.25cm} h^{\text{NLoS}}_{pq} (t) \hspace*{-0.225cm}&=&\hspace*{-0.225cm} \sum_{\ell \in L} \sum^{\ell_N}_{n = 1} e^{j \varphi_{\ell_n} -j \frac{2 \pi}{\lambda} \big(\xi_{T,\ell_n} + \xi_{R,\ell_n} \big) } \nonumber \\ [0.125cm]
\hspace*{-0.225cm} &\times& \hspace*{-0.225cm} e^{j \frac{2\pi}{\lambda} k_{p_h} \delta_T \cos (\alpha_{T,\ell_n} - \psi_T ) \cos\beta_{T,\ell_n} } \nonumber \\ [0.125cm]
\hspace*{-0.225cm} &\times& \hspace*{-0.225cm} e^{j \frac{2\pi}{\lambda} k_{p_v} \delta_T \sin\beta_{T,\ell_n} } \nonumber \\ [0.125cm]
\hspace*{-0.225cm} &\times& \hspace*{-0.225cm} e^{j \frac{2\pi}{\lambda} k_q \delta_R \cos\big(\alpha_{R,\ell_n} (t) - \psi_R \big) \cos\beta_{R,\ell_n} (t) \cos{\theta_R}} \nonumber \\ [0.125cm]
\hspace*{-0.225cm} &\times& \hspace*{-0.225cm} e^{j \frac{2\pi}{\lambda} k_q \delta_R \sin\beta_{R,\ell_n} (t) \sin{\theta_R}} \nonumber \\ [0.125cm]
\hspace*{-0.225cm}&\times&\hspace*{-0.225cm}  e^{j \frac{2\pi}{\lambda} v_R t \cos \big( \alpha_{R,\ell_n}(t) - \eta_R \big) \cos{\beta_{R,\ell_n} (t)} } \, ,
\end{eqnarray}
with $\{\varphi_{\ell_n}\}_{n = 1,2,...,L_n}$ being the independent and uniformly distributed random phase, i.e., $\varphi_{\ell_n} \sim \text{U}[-\pi,\pi)$. The $\alpha_{T,\ell_n} (t)$ and $\beta_{T,\ell_n} (t)$ are respectively the angles of departure of the transmitted waves in the azimuth and vertical planes, which correspond to the midpoint of $(p^{\text{sub}}_h, p^{\text{sub}}_v)$-th subarray to the $n$-th path in the $\ell$-th cluster. They are respectively derived by
\begin{eqnarray}
\hspace*{-1.5cm} \alpha_{T,\ell_n} \hspace*{-0.225cm}&=&\hspace*{-0.225cm} \arctan\frac{y_{\ell_n} - d_{(p^{\text{sub}}_{h,v}),y}}{x_{\ell_n} - d_{(p^{\text{sub}}_{h,v}),x}} \, ,
\end{eqnarray}
\begin{eqnarray}
\hspace*{-1.5cm} \beta_{T,\ell_n}  \hspace*{-0.225cm}&=&\hspace*{-0.225cm} \nonumber\\
&&\hspace*{-1.35cm} \arctan\frac{z_{\ell_n} - d_{(p^{\text{sub}}_{h,v}),z}}{\sqrt{ (x_{\ell_n} - d_{(p^{\text{sub}}_{h,v}),x})^2 + (y_{\ell_n} - d_{(p^{\text{sub}}_{h,v}),y})^2 }} \, .
\end{eqnarray}
Similarly, $\alpha_{R,\ell_n} (t)$ and $\beta_{R,\ell_n} (t)$ are respectively the angles of arrival of the waves in the azimuth and vertical planes, which is derived by calculating the angle between the line connecting the $q$-th element in MR antenna array and the $n$-th path in the $\ell$-th cluster and the $x$ axis, thereby can be shown as follows:
\begin{eqnarray}
\alpha_{R,\ell_n} (t) \hspace*{-0.225cm}&=&\hspace*{-0.225cm} \arctan\frac{y_{\ell_n} - d_{q,y}(t)}{x_{\ell_n} - d_{q,x}(t)} \, ,
\end{eqnarray}
\begin{eqnarray}
\beta_{R,\ell_n} (t) \hspace*{-0.225cm}&=&\hspace*{-0.225cm} \arctan\frac{z_{\ell_n} - d_{q,z}}{\sqrt{ (x_{\ell_n} - d_{q,x}(t) )^2 + (y_{\ell_n} - d_{q,y}(t))^2 }}. \, \nonumber \\
\end{eqnarray}
In order to describe the sparse features of the scatterers in the large-scale MIMO communication channel model, the von Mises probability density function (PDF) is adopted to generate the angle parameters, which can be shown as follows:
\begin{eqnarray}
f( \alpha ) = \frac{e^{\kappa \cos{(\alpha - \mu_{\alpha})}}} {2\pi I_0(\kappa)}
\end{eqnarray}
with $\kappa$ being the environmental factor \cite{JiangRIS2023}. The $\mu_{\alpha}$ stands for the mean value of $\alpha$, and $I_0(\cdot)$ represents a modified Bessel function of order $0$. By utilizing the azimuth and vertical angles of departure generated by the von Mises distribution, we can accurately calculate the position of the scatterers. This approach is helpful to effectively model the distribution of scatterers within the presented communication model, thereby enhancing our modeling precision. {\color{blue} It can be seen that the channel coefficients for the NLoS and LoS propagation links are mainly composed of the phase, distance parameters, antenna steering vectors at the BS and user sides, as well as doppler phase shift. This intuitively reflects the relationships between the  model parameters and the physical features of large-scale MIMO channels.}

Notice that the subarray decomposition scheme mainly applies to the sub-6G scenario. In fact, the propagation characteristics correspond to different frequency bands have obviously differences, such as the sparse mmWave channel model and environmentally sensitive terahertz communication. Hence, the designed subarray decomposition framework is not be suitable for other scenarios. Furthermore, our communication model is primarily employed to characterize the propagation environment of BS-to-MR, with its core objective being to assist system design rather than enhancing system performance. {\color{blue} Compared with the existing works, the channel model presented in this paper introduces the subarray decomposition framework, which aims at providing a practical communication algorithm with lower complexity.}

\section{Propagation Properties of the Proposed Channel Model}
In this part, we will derive and discuss the propagation properties of the channel model presented in this paper based upon the subarray decomposition scheme. The relevant numerical results will be presented and discussed in the next section.

\subsection{ST CCFs}

In the existing literature, the precision accuracy of a wireless channel is often validated through the derivation of the ST CCF. This index is measured by the correlation derivations between two distinct complex CIRs, i.e., $h_{pq} (t)$ and $h_{p'q'} (t + \Delta t)$ with $p'$ locating at the $p'_h$-th $(p'_h = 1,2,...,P_h)$ row and $p'_v$-th $(p'_v = 1,2,...,P_v)$ column of the transmitter array and $q' = 1,2,...,Q$. It can be derived by the following equation:
\begin{eqnarray}\label{eq:rho}
\rho_{h_{pq} h_{p'q'}} (t, \Delta_p, \Delta_q, \Delta t) = \mathbb{E} \Big[ \frac{h_{pq} (t) h^*_{p'q'} (t + \Delta t) } {\vert h_{pq} (t) \vert \vert h^*_{p'q'} (t + \Delta t) \vert} \Big] \, ,
\end{eqnarray}
where $\Delta_p$ denotes the normalized antenna spacing from the $p$-th antenna unit to the $p'$-th antenna unit at the BS side, and $\Delta_q$ is that from the $q$-th antenna unit to the $q'$-th antenna unit at the MR side. By combining \eqref{eq:h_pq}, the ST CCF of the proposed communication model can be derived by
\begin{eqnarray}\label{eq:rho_LoS_NLoS}
\hspace*{-1cm} \rho_{h_{pq} h_{p'q'}} (t, \Delta_p, \Delta_q, \Delta t) \hspace*{-0.225cm}&=&\hspace*{-0.225cm} \rho^{\text{LoS}}_{h_{pq} h_{p'q'}} (t, \Delta_p, \Delta_q, \Delta t) \nonumber \\[0.125cm]
\hspace*{-0.225cm}&+&\hspace*{-0.225cm} \rho^{\text{NLoS}}_{h_{pq} h_{p'q'}} (t, \Delta_p, \Delta_q, \Delta t) \, ,
\end{eqnarray}
where $\rho^{\text{LoS}}_{h_{pq} h_{p'q'}} (t, \Delta_p, \Delta_q, \Delta t)$ stands for the ST CCF of the LoS propagation links, which can be derived by substituting \eqref{eq:h_LoS} into \eqref{eq:rho}, thus we have
\allowdisplaybreaks[3]
\begin{eqnarray}\label{eq:rho_LoS}
\hspace*{-0.5cm}\rho^{\text{LoS}}_{h_{pq} h_{p'q'}} (t, \Delta_p, \Delta_q, \Delta t) \hspace*{-0.225cm}&=&\hspace*{-0.225cm} \frac{K}{K+1} \nonumber \\[0.125cm]
&&\hspace*{-3.5cm} \times \hspace*{0.05cm} e^{-j \frac{2 \pi}{\lambda} \big( \xi_{T,R} (t) - \xi_{T,R} (t + \Delta t) \big) } \nonumber \\[0.125cm]
&&\hspace*{-3.5cm} \times \hspace*{0.05cm} e^{j \frac{2\pi}{\lambda} k_{p_h} \delta_T \cos \big(\alpha^{\text{LoS}}_{T}(t) - \psi_T \big) \cos\beta^{\text{LoS}}_{T}(t) } \nonumber \\[0.125cm]
&&\hspace*{-3.5cm} \times \hspace*{0.05cm} e^{j \frac{2\pi}{\lambda} k_{p_v} \delta_T \sin\beta^{\text{LoS}}_{T}(t) } \nonumber \\[0.125cm]
&&\hspace*{-3.5cm} \times \hspace*{0.05cm} e^{-j \frac{2\pi}{\lambda} k_{p'_h} \delta_T \cos \big(\alpha^{\text{LoS}}_{T}(t + \Delta t) - \psi_T \big) \cos\beta^{\text{LoS}}_{T}(t + \Delta t) } \nonumber \\[0.125cm]
&&\hspace*{-3.5cm} \times \hspace*{0.05cm} e^{-j \frac{2\pi}{\lambda} k_{p'_v} \delta_T \sin\beta^{\text{LoS}}_{T}(t + \Delta t) } \nonumber \\[0.125cm]
&&\hspace*{-3.5cm} \times \hspace*{0.05cm} e^{j \frac{2\pi}{\lambda} k_q \delta_R \cos \big(\alpha^{\text{LoS}}_{R}(t) - \psi_R \big) \cos\beta^{\text{LoS}}_{R}(t) \cos{\theta_R} } \nonumber \\[0.125cm]
&&\hspace*{-3.5cm} \times \hspace*{0.05cm} e^{j \frac{2\pi}{\lambda} k_q \delta_R \sin\beta^{\text{LoS}}_{R}(t) \sin{\theta_R} } \nonumber \\[0.125cm]
&&\hspace*{-3.5cm} \times \hspace*{0.05cm} e^{-j \frac{2\pi}{\lambda} k_{q'} \delta_R \cos \big(\alpha^{\text{LoS}}_{R}(t+ \Delta t) - \psi_R \big) \cos\beta^{\text{LoS}}_{R}(t+ \Delta t) \cos{\theta_R} } \nonumber \\[0.125cm]
&&\hspace*{-3.5cm} \times \hspace*{0.05cm} e^{-j \frac{2\pi}{\lambda} k_{q'} \delta_R \sin\beta^{\text{LoS}}_{R}(t+ \Delta t) \sin{\theta_R} } \nonumber \\[0.125cm]
&&\hspace*{-3.5cm} \times \hspace*{0.05cm} e^{j \frac{2\pi}{\lambda} v_R t \cos \big(\alpha^{\text{LoS}}_{R}(t) - \eta_R \big) \cos{\beta^{\text{LoS}}_{R}(t)} } \nonumber \\[0.125cm]
&&\hspace*{-3.5cm} \times \hspace*{0.05cm} e^{-j \frac{2\pi}{\lambda} v_R (t + \Delta t) \cos \big( \alpha^{\text{LoS}}_{R} (t + \Delta t) - \eta_R \big) \cos{\beta^{\text{LoS}}_{R}(t + \Delta t)} } \, .
\end{eqnarray}

In \eqref{eq:rho_LoS_NLoS}, $\rho^{\text{NLoS}}_{h_{pq} h_{p'q'}} (t, \Delta_p, \Delta_q, \Delta t)$ stands for the ST CCF of the LoS propagation link. By substituting \eqref{eq:h_NLoS} into \eqref{eq:rho}, we can derive the corresponding expression as
\allowdisplaybreaks[3]
\begin{eqnarray}\label{eq:rho_NLoS}
\hspace*{-0.2cm} \rho^{\text{NLoS}}_{h_{pq} h_{p'q'}} (t, \Delta_p, \Delta_q, \Delta t) \hspace*{-0.225cm} &=& \hspace*{-0.225cm} \sum_{\ell \in L} \sum^{\ell_N}_{n = 1} \frac{1}{K+1} \nonumber \\ [0.125cm]
&&\hspace*{-3.8cm} \times \hspace*{0.05cm} e^{-j \frac{2 \pi}{\lambda} \big(\xi_{R,\ell_n} (t) - \xi_{R,\ell_n} (t + \Delta t) \big) } \nonumber \\ [0.125cm]
&&\hspace*{-3.8cm} \times \hspace*{0.05cm} e^{j \frac{2\pi}{\lambda} k_q \delta_R \cos\big(\alpha_{R,\ell_n} (t) - \psi_R \big) \cos\beta_{R,\ell_n} (t) \cos{\theta_R} } \nonumber \\ [0.125cm]
&&\hspace*{-3.8cm} \times \hspace*{0.05cm} e^{j \frac{2\pi}{\lambda} k_q \delta_R \sin\beta_{R,\ell_n} (t) \sin{\theta_R}} \nonumber \\ [0.125cm]
&&\hspace*{-3.8cm} \times \hspace*{0.05cm} e^{-j \frac{2\pi}{\lambda} k_{q'} \delta_R \cos\big(\alpha_{R,\ell_n} (t + \Delta t) - \psi_R \big) \cos\beta_{R,\ell_n} (t + \Delta t) \cos{\theta_R} } \nonumber \\ [0.125cm]
&&\hspace*{-3.8cm} \times \hspace*{0.05cm} e^{-j \frac{2\pi}{\lambda} k_{q'} \delta_R \sin\beta_{R,\ell_n} (t + \Delta t) \sin{\theta_R}} \nonumber \\ [0.125cm]
&&\hspace*{-3.8cm} \times \hspace*{0.05cm}  e^{j \frac{2\pi}{\lambda} v_R t \cos \big( \alpha_{R,\ell_n}(t) - \eta_R \big) \cos{\beta_{R,\ell_n} (t)} } \nonumber \\ [0.125cm]
&&\hspace*{-3.8cm} \times \hspace*{0.05cm}  e^{-j \frac{2\pi}{\lambda} v_R (t + \Delta t) \cos \big( \alpha_{R,\ell_n}(t + \Delta t) - \eta_R \big) \cos{\beta_{R,\ell_n} (t + \Delta t)} } \, .
\end{eqnarray}
It is found that the ST CCFs are affected by the positional relationship of the transmitter and the receiver, including the height of the transmitter, the motion direction and duration of the receiver.
Furthermore, upon applying the conditions $\Delta_p = \Delta_q = 0$, we can derive \eqref{eq:rho_LoS} and \eqref{eq:rho_NLoS} to represent the temporal ACFs of the communication model.

\subsection{Frequency CFs}

The frequency CF of the proposed subarray decomposition scheme based channel model is derived by
\begin{eqnarray}\label{eq:FCF}
\rho_{h_{pq}} (t, \Delta f) = \mathbb{E} \Big[ \frac{h_{pq} (t, f) h^*_{pq} (t, f + \Delta f) } {\vert h_{pq} (t, f) \vert \vert h^*_{pq} (t, f + \Delta f) \vert} \Big] \, ,
\end{eqnarray}
where $\Delta f$ stands for the frequency difference. By substituting \eqref{eq:h_pq} into \eqref{eq:FCF}, the frequency CF of the presented communication model can be expressed as
\begin{eqnarray}
\hspace*{-1cm} \rho_{h_{pq}} (t, \Delta f) \hspace*{-0.225cm}&=&\hspace*{-0.225cm} \rho^{\text{LoS}}_{h_{pq}} (t, \Delta f) + \rho^{\text{NLoS}}_{h_{pq}} (t, \Delta f) \, ,
\end{eqnarray}
where $ \rho^{\text{LoS}}_{h_{pq}} (t, \Delta f)$ and $\rho^{\text{NLoS}}_{h_{pq}} (t, \Delta f)$ are respectively the frequency CFs of the LoS and NLoS propagation links, which is derived by substituting \eqref{eq:h_LoS} and \eqref{eq:h_NLoS} into \eqref{eq:FCF}. It is worthy to note that the frequency CF exhibits a correlation with the motion time $t$, which reveals the non-stationarity of the presented communication model in time domain. In addition, both the height of the transmitter and the motion speed of the receiver contribute to the variation in the frequency CF of the presented communication model.

\subsection{Channel Capacities}

As a key index to evaluate the information transmission capability of large-scale MIMO systems, the channel capacity of the BS-to-MR communication system is derived by
\begin{eqnarray}\label{eq:CP}
C = \log_2 \Big( \det\big(\mathbf{I}_{Q} + \frac{\rho_{\text{SNR}}}{P} \overline{\mathbf{H}} (t,\tau) \overline{\mathbf{H}}^{\mathrm{H}} (t, \tau) \big) \Big) \, \,
\end{eqnarray}
with $\mathbf{I}_{Q}$ and $\rho_{\text{SNR}}$ being the identity matrix of size $Q \times Q$ and the signal-to-noise ratio (SNR), respectively. $\overline{\mathbf{H}} (t,\tau)$ denotes the normalized channel matrix, which is written as follows:
\begin{eqnarray}
\overline{\mathbf{H}} (t,\tau) = \mathbf{H} (t,\tau) \cdot \Big\{ \frac{1}{PQ} \Vert \mathbf{H} (t,\tau) \Vert_\text{F} \Big\}^{-\frac{1}{2}} \, ,
\end{eqnarray}
where $\Vert \cdot \Vert_\text{F}$ represents the Frobenius norm of a matrix.
It is crucial to note that the channel capacity is inherently influenced by multiple factors, including the number of antenna units at both the BS and MR sides. In addition, the positions of the BS and MR also play important roles in shaping the channel capacity.

\section{Results and Discussions}

\subsection{Simulation Setup}

In this section, numerical results on the propagation characteristics of the proposed communication model are conducted.
{\color{blue}
The simulation parameters for characterizing the propagation features are set as follows: $f_c = 5$ GHz, $H_0 = 20$ m, $D_0 = 50$ m, $P_h = P_v = 64$, $Q = 4$, $\delta_T = \delta_R = \lambda/2$, $\psi_T = \psi_R = \pi/2$, $\theta_R = \pi/3$, $v_R = 5$ m/s, and $\eta_R = \pi/2$. In this case, the Rayleigh distance of the large-scale MIMO antenna array is approximately 267 m, which can ensure that the terminal is within the near-field region of the BS.}

\subsection{Channel Modeling Precision and Complexity}

Existing literatures have widely regarded the spherical wavefront model as the mainstream method to achieve the high precision for investigating wireless propagation characteristics. Consequently, by calculating the error between the proposed model and the spherical wavefront model, we represent the normalized absolute error $\Delta$ to evaluate the performance of modeling precision. We have
\begin{align}\label{eq:Delta}
\Delta = 10 \log_{10} \Big\{\sum^{P_h}_{p_h = 1} \sum^{P_v}_{p_v = 1} \sum^{Q}_{q = 1} \frac{\vert h_{pq} (t, \tau) - h^{\text{spherical}}_{pq} (t, \tau) \vert}{\vert h^{\text{spherical}}_{pq} (t, \tau) \vert} \Big\} \, ,
\end{align}
where $h^{\text{spherical}}_{pq} (t, \tau)$ stands for the complex CIR of the spherical wavefront model. Here, a higher value of $\Delta$ indicates a greater deviation between the model and the spherical wavefront model, signifying a larger error.

\begin{figure}[!t]
  \centering
  \includegraphics[width=8.5cm]{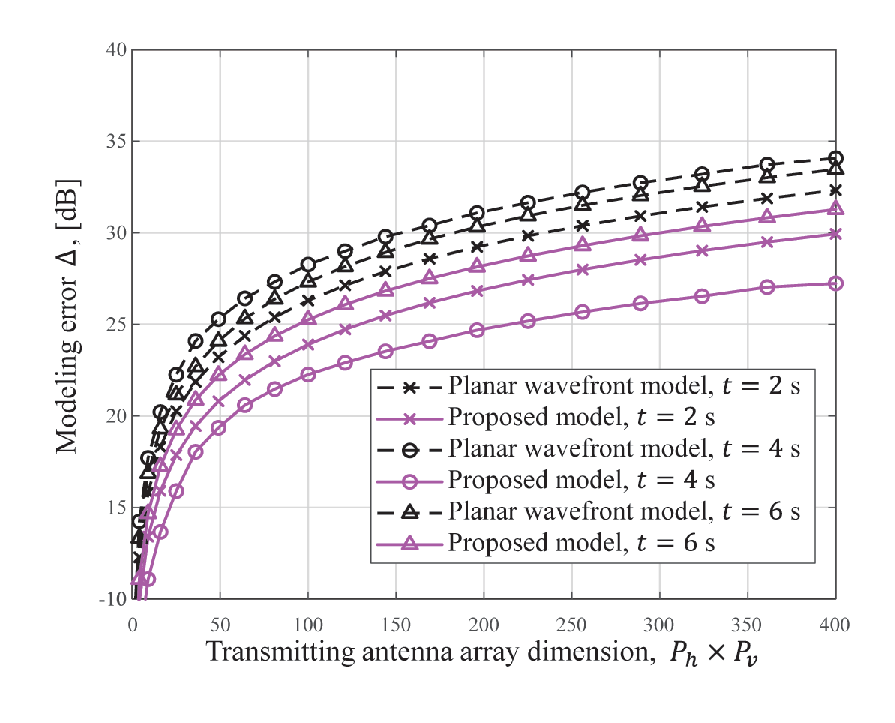}\\
  \caption{Comparisons between the modeling error performances based upon the subarray decomposition scheme and those based upon the planar wavefront model.}
  \label{fig3}
\end{figure}

By utilizing \eqref{eq:Delta}, Fig. \ref{fig3} reveals the modeling precision performance of the large-scale MIMO channels based upon the subarray decomposition scheme. It can be observed that the modeling error continually increases with the rising of the dimensions of the antenna apertures, which is in accordance with the results in \cite{Xiong2023}. As comparison to the conventional channel model that relies on planar wave assumptions, the employment of the subarray decomposition scheme significantly enhances the performance of the proposed channel modeling framework. Especially for the case of large antenna arrays, which are typically associated with the near-field region. Another phenomenon is that when the number of MIMO antenna arrays at BS side is small, the modeling precision is basically equivalent to that of the planar wavefront model. However, as the number of antennas continues to increase, the precision advantage of the presented communication model becomes more obviously. This observation underscores the capability of the designed subarray decomposition framework in efficiently describing the performance of large-scale MIMO communication systems. In addition, we can observe that the channel modeling error behave differently at distinct instants. This phenomenon is caused by the temporal variations in the distance between the BS and MR due to the motion of the MR, which is also confirmed by the observations of \cite{Jiang2023}.

\begin{figure}
  \centering
  \includegraphics[width=8.5cm]{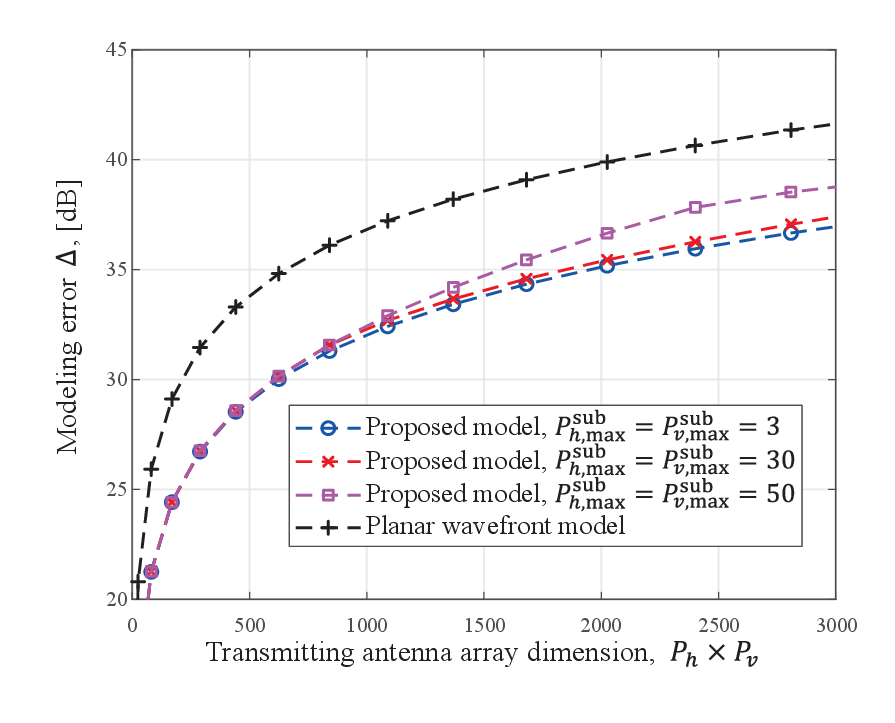}\\
  \caption{Comparisons between the modeling error performances based upon the subarray decomposition scheme regarding the different decomposing methods.}
  \label{fig4}
\end{figure}

Fig. \ref{fig4} depicts the modeling precision of the large-scale MIMO channels in compliance with the subarray decomposition scheme under different decomposing methods. It should be noted that upon setting the size of the largest subarray to $30 \times 30$, i.e., $P^{\text{sub}}_{h,\text{max}} = P^{\text{sub}}_{v,\text{max}} = 30$, the MR is approximately at the boundary of the near- and far-field ranges. We observe that when the MR is in the far-field scenario, the modeling precision is basically unchanged regardless of how the subarray is decomposed. This is caused by the fact that the subarray decomposition scheme have limitations to adapt far-field conditions. Furthermore, when the MR is in the far-field range, a smaller size of the largest subarray means that the entire channel is superimposed by more subarrays of planar wavefront, which results in lower modeling error.

To evaluate the modeling complexity of the designed framework, we introduce the concept of ``real operations (ROs)'' defined in \cite{Zhang2019}. The implementation of any mathematical operation is based upon four basic operations: addition, multiplication, division of real numbers, and lookup tables, each of these operations cost one RO. Since the complex CIR of the presented communication model coincides with that of the traditional geometry-based model, we herein only focus on the numbers of ROs required to generate the angle parameters, which can be expressed as
\begin{align}\label{eq:C}
C \approx P_h P_v Q (C_{\text{LoS}} + C_{\text{NLoS}}) \, ,
\end{align}
where $C_{\text{LoS}}$ and $C_{\text{NLoS}}$ represent the average numbers of ROs for each angle calculation in LoS and NLoS propagation links, respectively. Note that each subarray only needs to calculate the angle parameter once. Therefore, $C_{\text{LoS}}$ and $C_{\text{NLoS}}$ can be respectively expressed as
\begin{align}\label{eq:C_LoS}
C_{\text{LoS}} = \frac{C^{\text{sub}}_{\text{LoS}} P^{\text{sub}}_h P^{\text{sub}}_v}{P_h P_v} \, ,
\end{align}
\begin{align}\label{eq:C_NLoS}
C_{\text{NLoS}} = \frac{C^{\text{sub}}_{\text{NLoS}} P^{\text{sub}}_h P^{\text{sub}}_v}{P_h P_v} \, ,
\end{align}
with $C^{\text{sub}}_{\text{LoS}}$ and $C^{\text{sub}}_{\text{NLoS}}$ denoting the numbers of ROs required for each angle calculation for subarray in LoS and NLoS propagation links, respectively. By substituting \eqref{eq:C_LoS} and \eqref{eq:C_NLoS} into \eqref{eq:C}, we can rewrite \eqref{eq:C} as follows:
\begin{align}
C \approx P^{\text{sub}}_h P^{\text{sub}}_v Q (C^{\text{sub}}_{\text{LoS}} + C^{\text{sub}}_{\text{NLoS}}) \, ,
\end{align}
where $C^{\text{sub}}_{\text{LoS}}$ and $C^{\text{sub}}_{\text{NLoS}}$ are respectively the numbers of ROs required for each angle calculation for subarray in LoS and NLoS propagation links.

On the premise of simply comparing the computational complexity of angle parameters, according to (12) and (13), the required operations for LoS propagation link include $6$ addition or subtraction operations ($6$ ROs), $2$ division operations ($2$ ROs), $2$ square operations ($2$ ROs), $1$ square root operation ($1$ ROs), $2$ arctangent operations ($32$ ROs), and $4$ assignment operations ($4$ ROs), resulting in $C^{\text{sub}}_{\text{LoS}} = 6+2+2+1+32+4 = 47$ ROs. Similarly, the operations required for the NLoS propagation link is derived as $C^{\text{sub}}_{\text{NLoS}} = 10+4+4+2+64+4 = 88$ ROs.

\begin{figure}
  \centering
  \includegraphics[width=8.5cm]{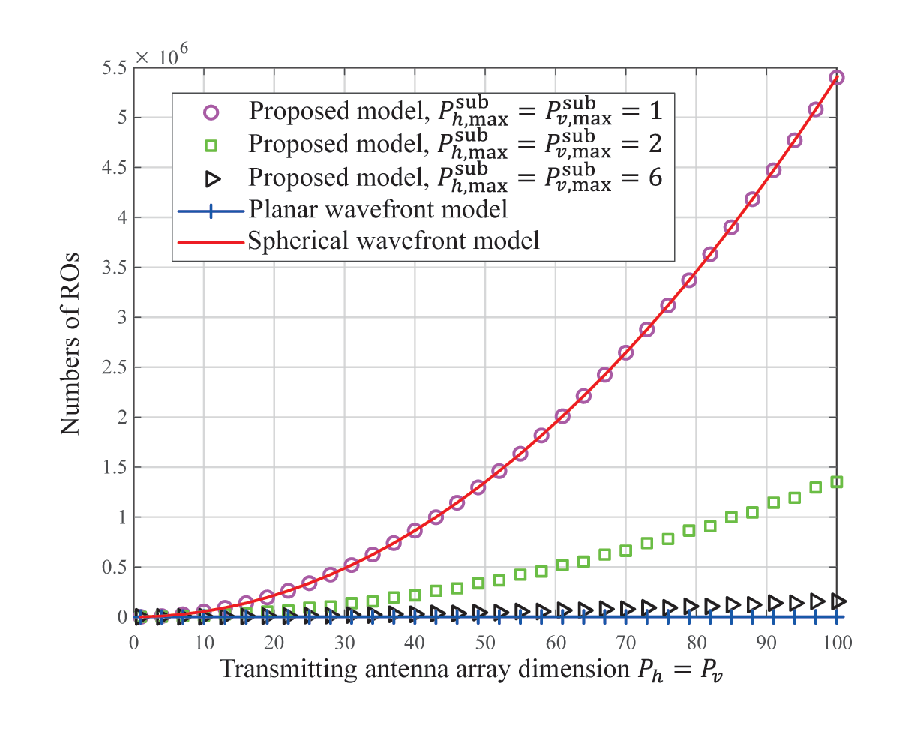}\\
  \caption{Comparisons between the complexities of the BS-MR channel model based upon the subarray decomposition scheme and those based upon the planar/spherical wavefront models.}
  \label{fig5}
\end{figure}

Fig. \ref{fig5} presents the modeling complexities of the BS-US communication channel in regard to the different methods. It is obvious that when we set size of the largest subarray as $1$, i.e., $P^{\text{sub}}_{h,\text{max}} = P^{\text{sub}}_{v,\text{max}} = 1$, the changing curves of the presented communication model coincide with that of the spherical wavefront model, which verifies the precision of the derivations of the designed subarray decomposition scheme. We also find that a larger size of the largest subarray results in a lower value of the modeling complexity, demonstrating that the BS-to-MR communication model based upon the subarray decomposition framework exhibits low computational complexity. Combined with the observations in Fig. \ref{fig4}, it can be concluded that the decomposition selection of the largest subarray needs to balance both modeling precision and computational complexity. Although the larger size of the largest subarray can greatly reduce the modeling complexity, its precision may be unacceptable. Conversely, the smaller size of the largest subarray will still maintain a high computational complexity, which indicates that the subarray needed to be decomposed in accordance with the actual communication requirements.

\subsection{Spatial CCFs}

\begin{figure}
  \centering
  \includegraphics[width=8.5cm]{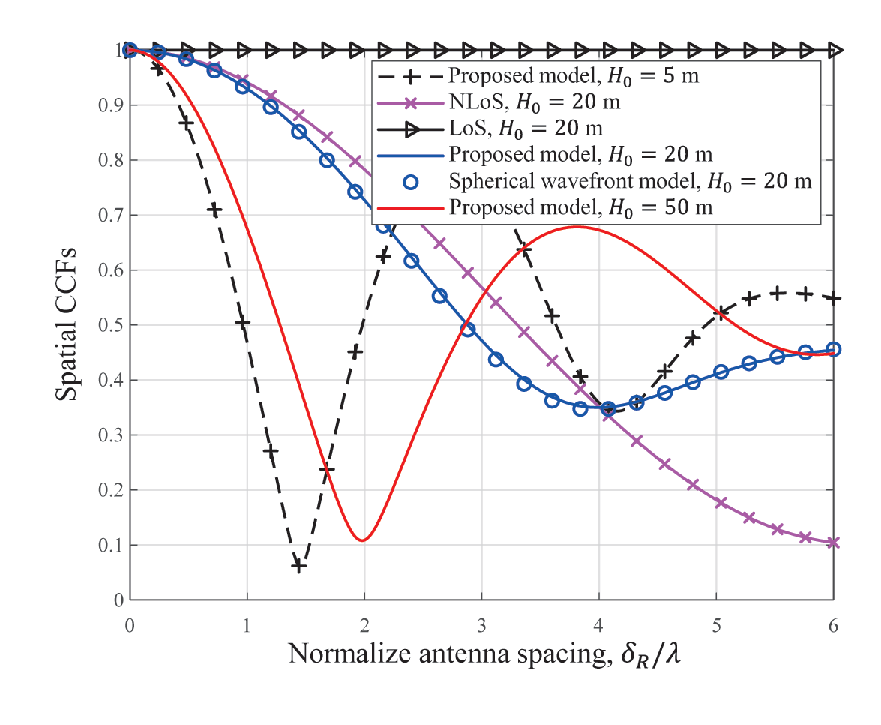}\\
  \caption{Spatial CCFs of the presented communication model regarding the different heights of the BS.}
  \label{fig6}
\end{figure}

Fig. \ref{fig6} illustrates the spatial CCFs of the presented communication model in regard to different heights of the BS. Obviously, the changing trends of the curves for spatial correlations are almost the same as those of the spherical wavefront model, which proves that the designed subarray decomposition framework has well modeling precision. In addition, we notice that when the height of the BS varies, the spatial CCFs exhibit distinct declining trends, which reflects the non-stationarity of the proposed channel model in spatial domain and it can also serve as a valuable reference for the practical deployment of MIMO systems.

\begin{figure}
  \centering
  \includegraphics[width=8.5cm]{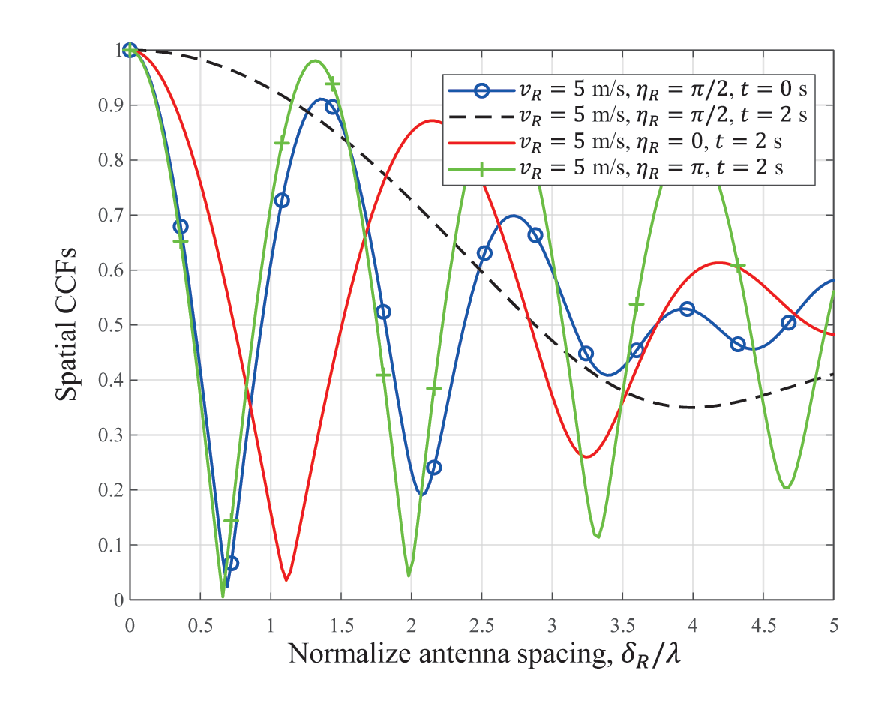}\\
  \caption{Spatial CCFs of the presented communication model regarding the different motion direction and duration of the receiver.}
  \label{fig7}
\end{figure}
In Fig. \ref{fig7}, we provide the spatial CCFs of the presented communication model based upon the subarray decomposition framework in regard to the different motion direction and duration of the MR. We find that the spatial correlation properties of the propagation links exhibit distinct decreasing trends on the motion direction of the receiver. Notably, the rate of decrease in spatial CCFs is more significant when the MR is closer to the BS. In addition, as time $t$ grows from 0 s to 2 s, the reduction rates of spatial CCFs observed within the presented communication model slow down correspondingly.

\subsection{Temporal ACFs}

\begin{figure}
  \centering
  \includegraphics[width=8.5cm]{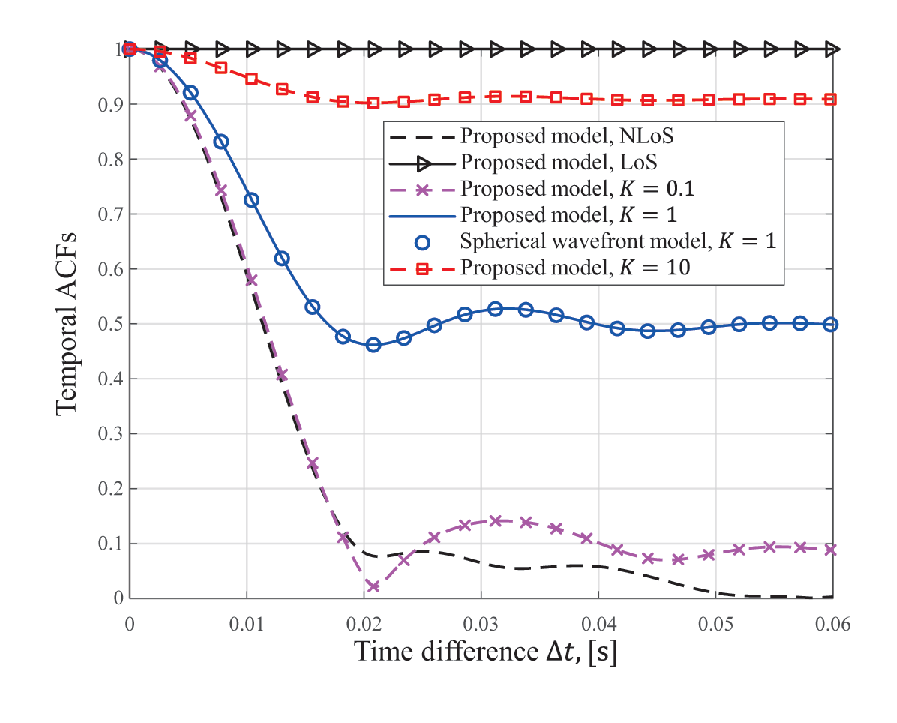}\\
  \caption{Temporal ACFs of the presented communication model regarding the different Rician factors.}
  \label{fig8}
\end{figure}

Fig. \ref{fig8} shows the temporal ACFs of the presented communication model in regard to different Rician factors. It is obvious that the changing curves of time correlation properties of the BS-MR channel model based upon the subarray decomposition framework closely align with those based upon the spherical wavefront model, which are similar to the observations in Fig. \ref{fig6}, thereby verifying the advantages of the designed subarray decomposition framework. Furthermore, the percentage of LoS components within the whole link configuration have great influences on the channel characteristics.
{\color{blue}
It is worth mentioning that this paper only focuses on the analysis of small-scale fading characteristics; therefore, we introduce the Rician factor $K$ for representing the proportional relationships between the NLoS propagation paths and LoS paths in the channel model. Specifically, when the value of $K$ is set to be a low value, i.e. $K=0.1$, which indicates that the proportion of NLoS propagation links to the LoS paths is extremely low, the proposed channel model approaches to be the Rayleigh channel. In this case, the temporal correlation is relatively low. However, when the $K$ gradually increase from $0.1$ to $10$, the proportion of the NLoS propagation paths to LoS paths would increase correspondingly, which lead to a phenomenon that the temporal correlations rise gradually. The above observations align with the simulated results in \cite{Mao2023} well, which validate the correctness of the derivations and conclusions.}

\begin{figure}
  \centering
  \includegraphics[width=8.5cm]{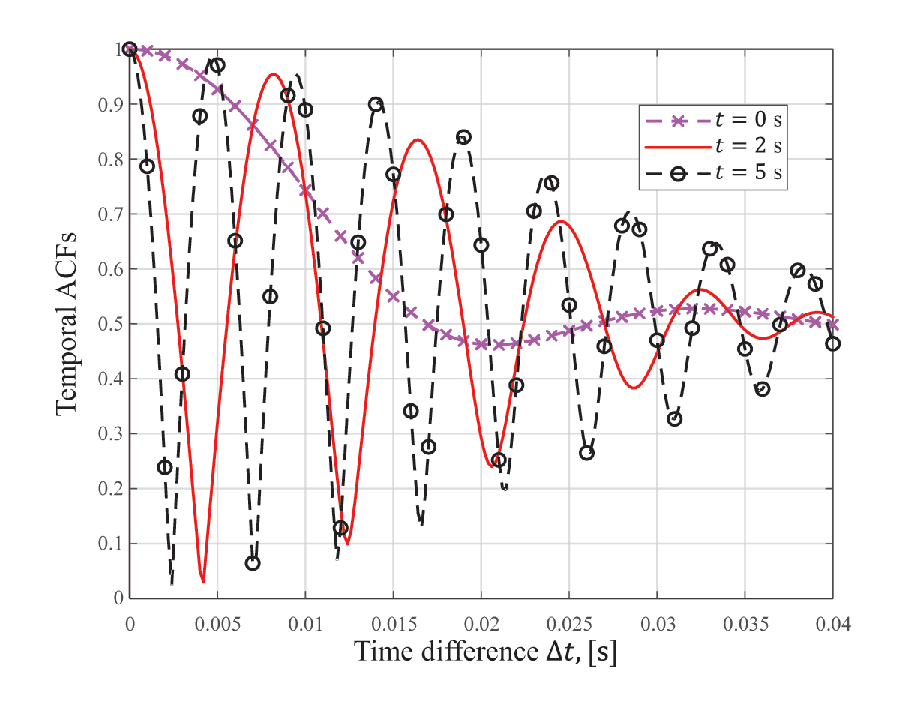}\\
  \caption{Temporal ACFs of the presented communication model regarding the different motion time of the receiver.}
  \label{fig9}
\end{figure}

Fig. \ref{fig9} illustrates the temporal ACFs of the presented communication model in regard to different motion time of the MR. Obviously, the changing curves of the temporal correlations of propagation links are affected by the motion time $t$, which confirm the channel time non-stationarity. In addition, when we increase the value of the time $t$ from $0$ to $5$ s, more slightly fluctuations of the temporal correlation properties can be observed in the figure. This changing trend aligns with the observations made in \cite{Lian2019, Jiang2019}, thereby confirming the precision of the derivations and simulations of the presented channel model.

\subsection{Frequency CFs}

\begin{figure}
  \centering
  \includegraphics[width=8.5cm]{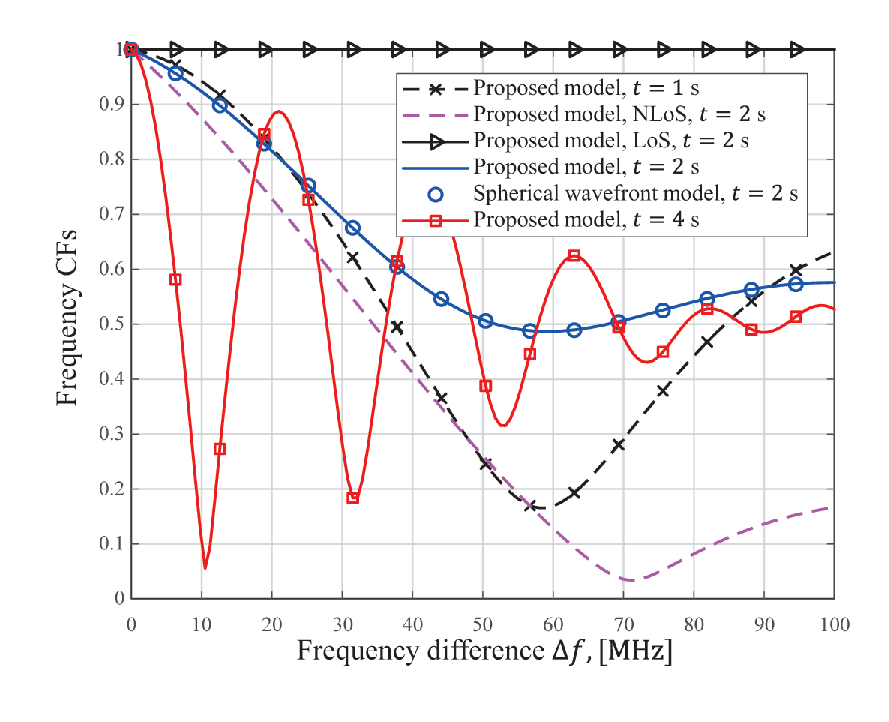}\\
  \caption{Normalized frequency CFs of the presented communication model regarding the different motion time.}
  \label{fig10}
\end{figure}

In Fig. \ref{fig10}, we study the influence of the motion time of the receiver on the channel frequency correlation properties. Evidently, a more substantial frequency difference $\Delta f$ leads to a lower in frequency correlation, highlighting the non-stationarity in frequency domain. Specifically, the fluctuating curves of the frequency correlations exhibit rapid variations in the initial stages, subsequently stabilizing as the frequency difference increases. The changing curve of the spherical wavefront model fits well with the presented communication model based upon the subarray decomposition framework, which confirms the observations in Figs. \ref{fig6} and \ref{fig8}. In addition, the frequency correlation curves of the presented communication model exhibit distinct trends with the changing of motion time.

\begin{figure}
  \centering
  \includegraphics[width=8.5cm]{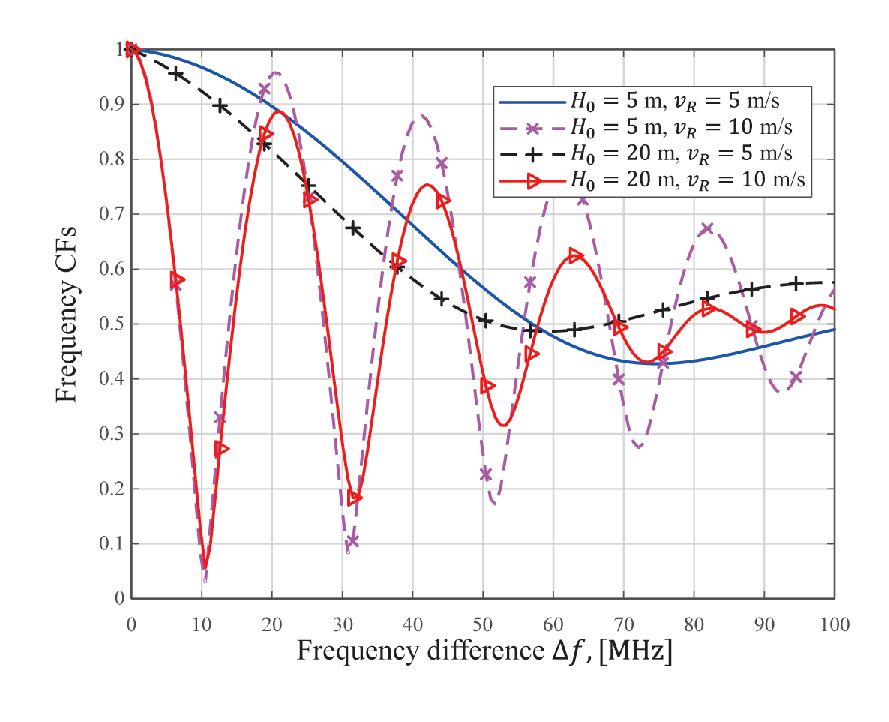}\\
  \caption{Normalized frequency CFs of the presented communication model regarding the different heights of the BS and different motion speeds of the receiver.}
  \label{fig11}
\end{figure}

Fig. \ref{fig11} presents the normalized frequency CFs of the presented communication model in regard to the different heights of the BS and different motion speeds of MR. It shows that the changing curves of the frequency correlation properties exhibit distinct declining trends as we set different values of the heights of the BS, aligning with the observations presented in Fig. \ref{fig6}. Furthermore, we can notice that the motion speed of the MR also has obvious impacts on the frequency correlation curve of the proposed model.

\subsection{Channel Capacities}

\begin{figure}
  \centering
  \includegraphics[width=8.5cm]{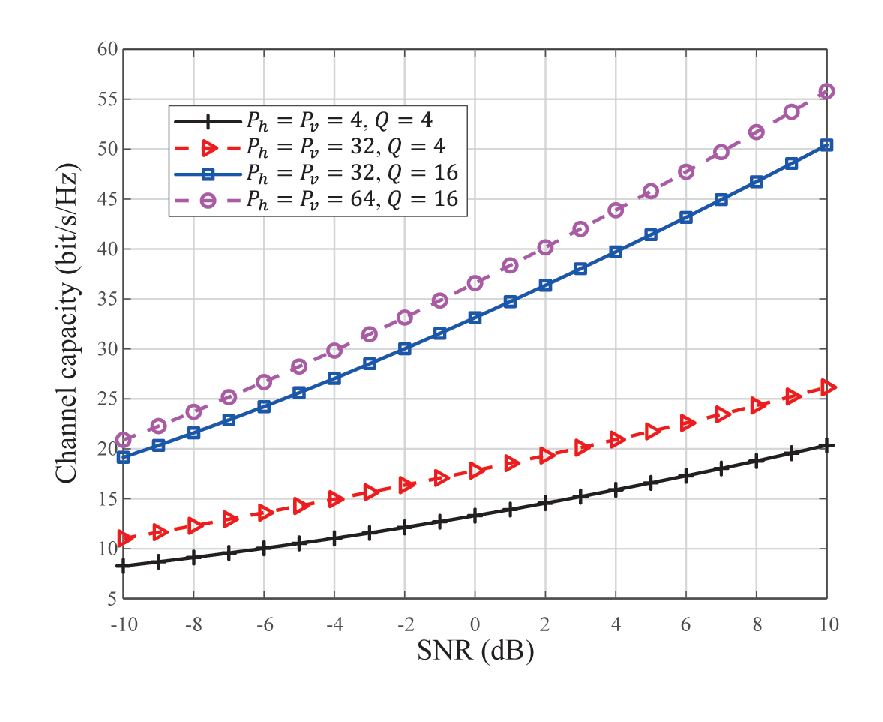}\\
  \caption{Channel capacities of the presented communication system regarding the different antenna configurations at the BS and the MR sides.}
  \label{fig12}
\end{figure}

By utilizing \eqref{eq:CP}, Fig. \ref{fig12} shows the channel capacities of the presented system in regard to the different antenna configurations at the BS and MR sides. It is obvious that the growth in the number of antenna units at the BS/MR side leads to a significant enhancement in the channel capacity, where same observations can be found in \cite{SunG}, which confirms the precision of the derivations of the channel capacities. Another phenomenon is that under the premise of $Q = 1$, when we increase the dimension of the transmitting antenna from $P_h = P_v = 32$ to $P_h = P_v = 64$, the magnitude of the increase in channel capacity decreases significantly. This is caused by the fact that the channel capacity tends to saturation, which provides valuable references for the system design of large-scale MIMO communications.

\begin{figure}
  \centering
  \includegraphics[width=8.5cm]{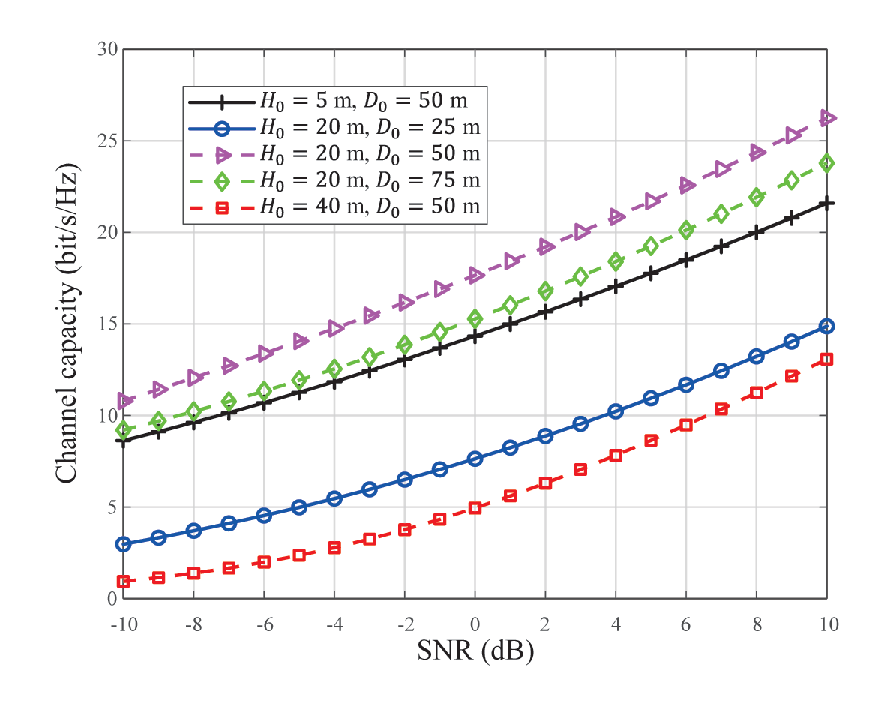}\\
  \caption{Channel capacities of the presented system regarding the different initial position parameters.}
  \label{fig13}
\end{figure}

In Fig. \ref{fig13}, we compare the channel capacities of the presented communication systems in regard to the different initial position parameters. Results indicate that channel capacity varies greatly for different heights of the BS. Specifically, as the height of the BS increases, the channel capacity initially exhibits an upward changing trend, subsequently reversing to a declining trend. Similarly, the channel capacity follows an analogous pattern with variations in the initial distance between the BS and the MR. These observations provide valuable insights into the antenna deployment and system design of the large-scale MIMO communications.



According to the Figs. \ref{fig3}-\ref{fig13}, we can conclude that the presented communication model based upon the subarray decomposition framework has the ability to reduce computational complexity by almost $75\%$, while maintaining high level of modeling precision. This observation obviously behave superior performance as compared with the conventional spherical wavefront model with high modeling complexity. As a result, the subarray decomposition scheme-based channel model is regarded as a viable solution for effectively balancing the modeling precision and complexity in large-scale MIMO communication systems.

\section{Conclusions}

In this paper, we have presented a 3D communication model for near-field large-scale MIMO communications between the BS and the MR. An efficient subarray decomposition scheme has been designed on the antenna array at the BS side, which has been confirmed to have the ability to balance the modeling precision and computational complexity. By employing this approach, we can effectively simulate the antenna interactions for large-scale MIMO communication systems in near-field scenarios, enabling better evaluation and optimization of the system performance. The numerical results have revealed that the channel characteristics of the presented communication model are affected by the height of the BS, the motion state of the MR, as well as the duration of its motion. Also, the antenna array configurations at the BS and MR sides have great influences on the channel capacity. Furthermore, the proposed scheme can reduce the complexity of modeling while ensuring a certain degree of precision accuracy.

{\color{blue}
As our future work, three promising directions can be presented: i) employ approximation algorithms to simplify the derivations of the complex CIRs, thereby further reducing the computational complexity of the proposed channel model; ii) undertake measurements for large-scale MIMO communications to further verify the numerical results of the propagation statistics based upon the designed subarray decomposition framework; iii) study large-scale MIMO communication systems in more complex conditions, such as hybrid near- and far-field scenarios.}

\end{document}